%% file: paper.tex
\documentclass[12pt]{article}

\newcommand{\MSOne}{\ensuremath{\normalfont\textsf{MS}_1}\xspace}
\newcommand{\MSTwo}{\ensuremath{\normalfont\textsf{MS}_2}\xspace}

\input{macros}

\title{Small-world networks and RNA secondary structures}
\author{Defne Surujon$^1$ \and Yann Ponty$^2$ \and Peter Clote$^1$\thanks{Corresponding author: {\tt clote@bc.edu}}}
\date{$1$: Biology Department, Boston College, USA.
Chestnut Hill, MA 02467\\
$2$: Laboratoire d'Informatiques, Ecole Polytechnique,
91128 Palaiseau Cedex - France.}

\begin{document}
\maketitle

\begin{abstract}
Let $\mathcal{S}_n$ denote the network of all RNA secondary structures of
length $n$, in which undirected edges exist between structures 
$s,t$ such that $t$ is obtained from $s$ by the addition, removal or shift 
of a single base pair. Using context-free grammars, generating functions
and complex analysis,
we show that the asymptotic average degree is $O(n)$ and that
the asymptotic clustering coefficient is $O(1/n)$, from which it follows that
the family $\mathcal{S}_n$, $n=1,2,3,\ldots$ of secondary structure networks 
is not small-world.
\end{abstract}

\section{Introduction}
\label{section:intro}

Small-world networks, first introduced in \cite{Watts.n98}, 
satisfy two properties:
(1) the {\em shortest path distance} between any two nodes is ``small''
(e.g.  six degrees of separation 
between any two persons), and
(2) the average {\em clustering coefficient} is large
(e.g. friends of a person tend to be friends of each other). 
Small-world networks appear to be ubiquitous in biology, sociology, and
information technology; indeed, examples include the neural network of 
{\em C. elegans} \cite{Watts.n98}, 
the gene co-expression in {\em S. cerevisiae} \cite{VanNoort.er04},
protein folding networks 
\cite{scalaProteinFoldingSmallWorld,Bowman.pnas10}, 
the low energy RNA secondary structure network 
of {\em E. coli} phe-tRNA \cite{Wuchty.nar03}, etc.
For additional examples, see the excellent review of
Albert and Barab{\'a}si \cite{barabasiReviewSmallWorld}.

In this paper, we investigate asymptotic properties of degree and
clustering coefficient for the ensemble of all RNA secondary structures, 
by using methods from algebraic combinatorics.  In particular, we rigorously 
prove that the network of RNA secondary structures is asymptotically
{\em not} small-world, although it displays strong differences from random 
networks.

\section{Preliminaries}
\label{section:basics}

\begin{figure}
\centering
\includegraphics[width=\textwidth]{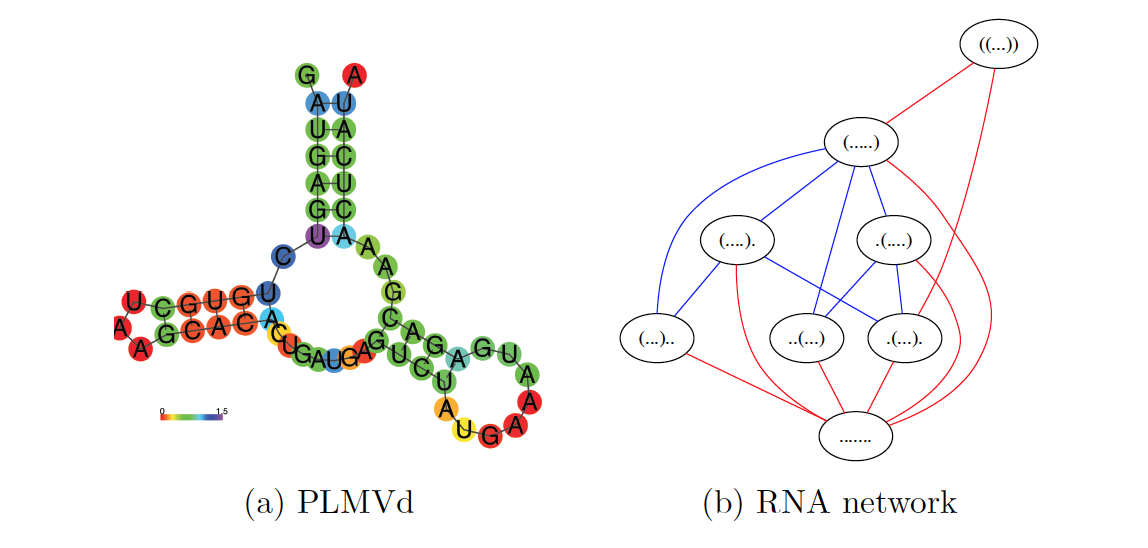}
\caption{{\em (a)} Consensus secondary structure of the type III 
hammerhead ribozyme from Peach Latent Mosaic Viroid (PLMVd) AJ005312.1/282-335
(isolate LS35, variant ls16b), taken from Rfam \cite{Gardner.nar11}
family RF00008. 
{\em (b)} 
Network for size 7 homopolymer with $\theta=3$, having 8 nodes and 
8 red $MS_1$ edges (base pair addition or removal), 8 blue $MS_2-MS_1$
edges (base pair shift), hence
a total of 16 $MS_2$ edges. It follows that $MS_1$ degree is 
$\frac{16}{8}=2$, while MS2 is $\frac{32}{8}=4$. 
}
\label{fig:secStr}
\end{figure}

In this section, we define notions of
RNA secondary structure, move sets $MS_1,MS_2$, and small-world networks.  
An RNA secondary structure of length $n$, subsequently called length $n$
structure, is defined to 
be a set $s$ of ordered pairs $(i,j)$, with $1 \leq i<j \leq n$, such that:
(1) There are no base triples; i.e.
if $(i,j),(k,\ell) \in s$ and $\{i,j\} \cap \{k,\ell\} \ne \emptyset$, 
then $i=k$ and $j=\ell$.
(2) There are no pseudoknots; i.e.
if $(i,j),(k,\ell) \in s$, then it is not the case that
$i<k<j<\ell$.
(3) There are at least $\theta=3$ unpaired bases in a hairpin loop; i.e.
if $(i,j) \in s$, then $j-i > \theta=3$. Note that base pairs
are {\em not} required to be Watson-Crick or wobble pairs, as is the case for
RNA molecules, such as that depicted in Figure~\ref{fig:secStr}a. 
This definition, sometimes called {\em homopolymer} secondary structure,
permits the  combinatorial analysis we employ to show that RNA networks
are not small-world.

Let $\mathcal{S}_n$ denote the set of all length $n$ structures.
The move sets $MS_1$ and $MS_2$, defined in \cite{flamm} for RNA secondary
folding kinetics, describe elementary moves that transform
a structure $s$ into another structure $t$.
Move set $MS_1$ [resp. $MS_2$] consists of either removing or adding 
[resp. removing, adding or shifting] a single base pair,
provided the resulting set of base pairs constitutes a valid 
structure, where shift moves are depicted in Figure~\ref{fig:shiftMoves}.
We overload the notation $\mathcal{S}_n$ to also denote the
$MS_1$ network [resp. $MS_2$ network], whose nodes are the length $n$
structures, where an undirected
edge between structures $s,t$ exists when $t$ is obtained
from $s$ by a single move from $MS_1$ [resp. $MS_2$]. 
Figure~\ref{fig:secStr}b shows
the $MS_1$ network (8 red edges) [resp. $MS_2$ network
(8 red and 8 blue edges)] for length $7$ structures,
where there are 8 nodes, $MS_1$ degree $\frac{16}{8}=2$ and $MS_2$ degree 
$\frac{32}{8}=4$.
See \cite{cloteJCC2015} and \cite{Clote.po15} for dynamic programming 
algorithms that
compute, respectively, the $MS_1$ and $MS_2$ degree for the network of
secondary structures of a given RNA sequence.

Small-world networks satisfy two conditions: (1) on average, 
the minimum path length between any two nodes is small, (2) neighbors
of a node tend to be connected to each other. The {\em global clustering
coefficient}, defined in equation~(77) of \cite{Newman.pres01}, is given by 
\begin{align}
\label{eqn:globalClusteringCoefficient}
\mathfrak{C}_{g}(G) 
&= \frac{3 \times \mbox{number of triangles}}{\mbox{number of connected triples}}
\end{align}
where a {\em triangle} is a set $\{ x,y,z \}$ of nodes, each of which
is connected by an edge, and a (connected) triple is a
set $\{ x,y,z \}$ of nodes, such that there is an edge from $x$ to $y$
and an edge from $x$ to $z$.  
Following \cite{contTanimura08}, the family
$\{ \mathcal{S}_n, n=1,2,3,\ldots \}$ of RNA networks is small-world if
the following conditions hold.
(1) There is a constant $c_1 \geq 0$, such that the minimum path length between
any two nodes of $\mathcal{S}_n$ is bounded above by $c_1 \ln n$.
(2) There is a constant $c_2 \geq 0$, such that the average network degree
of $\mathcal{S}_n$ is bounded above by $c_2 \ln n$.
(3) The global clustering coefficient is bounded away from zero.
By Theorem~\ref{thm:numSecStrMS1degree}, the network size of $\mathcal{S}_n$ 
is exponential in $n$. Since there are at most 
$\lfloor \frac{n-\theta}{2} \rfloor$ base pairs in any
length $n$ structure, the minimum path length between any two structures 
$s$ and $t$ is at most $n-\theta$; indeed, such a path can be achieved by 
successively removing each base pair $(i,j) \in s$, then successively
adding each base pair $(i,j)\in t$. It follows that
condition (1) is satisfied for both the $MS_1$ and
$MS_2$ networks of RNA structures. It is easy to see that the clustering
coefficient of the $MS_1$ network of RNA structures is zero, so in the 
remainder of the paper, we concentrate on conditions (2) and (3) for 
the $MS_2$ RNA network. 

Specific properties of RNA networks critically depend on the chosen 
definition of neighborhood of a structure, leading to the 
investigation of various move sets in the RNA kinetics literature. In
addition to the move sets $MS_1$ and $MS_2$ \cite{flamm}, which latter 
also models {\em defect diffusion} \cite{defectDiffusion}, 
other groups have considered more general move sets that allow
helix formation and 
disassociations \cite{Isambert.pnas00}.

The overall method used is as follows:
(1) Give a context-free grammar that 
generates the set of all secondary structures, possibly containing a
specific motif.
(2) Use Table~\ref{table:DSV} to derive and then solve
a functional relation for the complex generating function
$S(z)$, with the property that the $n$th Taylor coefficient of $S(z)$, 
denoted $[z^n]S(z)$, is equal to the number of length $n$ structures,
possibly containing a specific motif.
(3) Determine the dominant singularity and apply complex analysis
to obtain the asymptotic value of $[z^n]S(z)$.
For step (3), we use the Flajolet-Odlyzko Theorem,
stated as Corollary 2, part (i) on page 224 of \cite{FlaOdl90}. 
Before stating the theorem, 
we define the {\em dominant singularity} of  complex function
$f(z)$ to be the complex number $\rho$ having smallest absolute value 
(or modulus) at which $f(z)$ is not differentiable. 
\begin{theorem}[Flajolet  and Odlyzko]
\label{thm:flajolet}
Assume that $f(z)$ has a dominant singularity at $z=\rho>0$, is analytic 
for $z\ne \rho$ satisfying $|z| \leq |\rho|$, and that 
\begin{equation}
\label{alg-sing}
\lim\limits_{z \rightarrow \rho} f(z) = K(1-z/\rho)^{\alpha}.
\end{equation}
Then, as $n \rightarrow \infty$, if $\alpha \notin {0, 1, 2, ...}$,
\begin{eqnarray*} 
f_n = [z^n]f(z) \sim \frac{K}{\Gamma(-\alpha)} \cdot n^{-\alpha-1}
\cdot \rho^{-n}
\end{eqnarray*}
where $\sim$ denotes asymptotic equality and
$\Gamma$ denotes the Gamma function.
\end{theorem}
The plan of the paper is now as follows.
In Section~\ref{section:networkDegree}, we show that the average 
$MS_2$ degree of $\mathcal{S}_n$ is $O(n)$. In Section~\ref{section:triangles}
[resp. \ref{section:triples}] we prove that the average number of
triangles [resp. triples] is $O(n)$ [resp. $O(n^2)$], which
implies that the asymptotic global clustering coefficient is 
$O(1/n)$, hence not bounded away from zero. It follows
that the family of RNA secondary structure networks is not small-world.
Section~\ref{section:discussion} concludes the paper.
Due to space constraints, an appendix contains supplementary 
Figures S1-S6, a listing of the full grammar, and short explanation
of how to count all motifs necessary for an 
exact count of the global clustering coefficient. A
Mathematica\texttrademark program to compute the asymptotic global
clustering coefficient of $20.7728/n$ is available upon request.

\section{Expected network degree}
\label{section:networkDegree}
Due to space constraints, details for the computation 
of the asymptotic number of secondary structures as well as for 
$MS_1$ expected degree for homopolymers cannot be given here; however,
these results can be found in \cite{cloteArxivMS1degree}, including the
following.
\begin{theorem}
\label{thm:numSecStrMS1degree}
If $S(z)$ is the generating function for the number of secondary structures
for a homopolymer, then
\begin{align*}
[z^n] S(z) &\sim 0.713121 \cdot n^{-3/2} \cdot 2.28879^n
\end{align*}
If $MS_1\mbox{degree}(n)$ denotes the $MS_1$ expected network degree for
a homopolymer, then
\begin{align*}
MS_1\mbox{degree}(n) &\sim 0.473475 \cdot n
\end{align*}
\end{theorem}

\begin{figure}
\centering
\includegraphics[width=\textwidth]{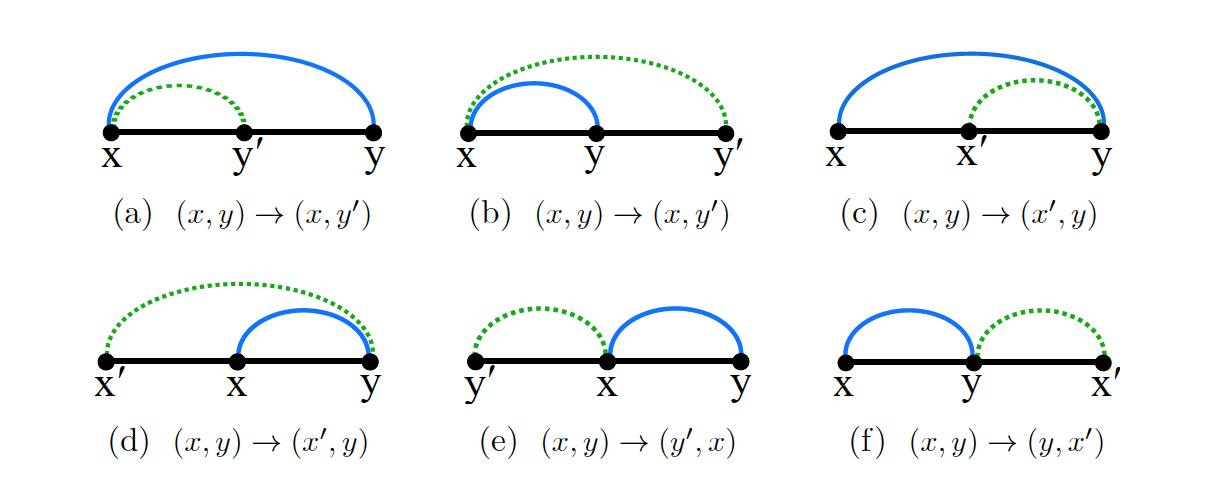}
\caption{Illustration of possible shift moves, where each subcaption
indicates the terminal symbols involved in the corresponding production
rule.  Solid blue [resp. dashed green] lines indicate the initial 
[resp. final] base pair in a shift move.
}
\label{fig:shiftMoves}
\end{figure}

Define the grammar $G$ to consist of the
terminal symbols $\op,\bullet,\cp,\oa,\star,\ca$, 
nonterminal symbols $\widehat{S},\widehat{T},S,R,\theta$, with start
symbol $\widehat{S}$.
Shift moves are represented in the grammar by one of the
three expressions: $\star \ca \ca$, $\oa \oa \star$, $\oa \star \ca$,
as depicted in Figure~\ref{fig:shiftMoves}. In particular,
$\star \ca \ca$ represents the right shift depicted
in Figure~\ref{fig:shiftMoves}a (ignoring possible intervening
structure), where base
pair $(x,y)$ is transformed to $(x,y')$ for $x<y'<y$; alternatively, the
$\star \ca \ca$ can represent the shift $(x,y)$ to $(x,y')$ for $x<y<y'$,
as depicted in Figure~\ref{fig:shiftMoves}b.
The expression $\oa \oa \star$ can represent the left shift depicted
in Figure~\ref{fig:shiftMoves}c, where base
pair $(x,y)$ is transformed to $(x',y)$ for $x<x'<y$; alternatively,
$\oa \oa \star$ can represent the shift $(x,y)$ to $(x',y)$ for $x'<x<y$,
as depicted in Figure~\ref{fig:shiftMoves}d.
The expression $\oa \star \ca$ can represent the right-to-left shift depicted
in Figure~\ref{fig:shiftMoves}e, where base
pair $(x,y)$ is transformed to $(y',x)$ for $y'<x<y$; alternatively,
$\oa \star \ca$
can represent the shift $(x,y)$ to $(y,x')$ for $x<y<x'$,
as depicted in Figure~\ref{fig:shiftMoves}f. The grammar
$G$ allows us to count the number of secondary structures, that
additionally contain a unique occurrence of exactly
one of the three expressions:
$\star \ca \ca$, $\oa \oa \star$, $\oa \star \ca$. 
The production rules of grammar $G$ are 
as follows:
\begin{align}
\label{eqn:grammarMS2theta3}
\widehat{S} &\rightarrow
\widehat{S} \bullet \,|\, \op \widehat{S} \cp \,|\,
S \op \widehat{S} \cp \,|\, \widehat{S} \op R \cp  \,|\, \widehat{T} 
\nonumber \\
\widehat{T} &\rightarrow 
\star R \ca \ca \,|\, S \star R \ca \ca \,|\, 
\star R \ca S \ca \,|\, S \star R \ca S \ca \,|\, \nonumber \\
& \oa \oa R \star \,|\, S \oa \oa R \star \,|\, 
\oa S \oa R \star \,|\, S \oa S \oa R \star \,|\, \nonumber \\
& \oa R \star R \ca \,|\, S \oa R \star R \ca \nonumber \\
S &\rightarrow \bullet \,|\, S \bullet \,|\, \op R \cp  \,|\, S \op R \cp 
\nonumber \\
R &\rightarrow \theta  \,|\, R \bullet \,|\, \op R \cp  \,|\, S \op R \cp 
\nonumber \\
\theta &\rightarrow \bullet \bullet \bullet 
\end{align}
The nonterminal $S$ is responsible for generating all
secondary structures of length greater than or equal to $1$.
In contrast, the
nonterminal $\widehat{S}$ is responsible for generating all
well-balanced expressions of length greater than or equal to $1$, that
involve exactly one of the three expressions:
$\star \ca \ca$, $\oa \oa \star$, $\oa \star \ca$.
To that end, the nonterminal $\widehat{T}$ is responsible
for generating all such
expressions, in which the rightmost symbol is either $\ca$ or $\star$,
but not $\bullet$ or $\cp$.
By induction on length of sequence generated, one can show that $G$
is a nonambiguous context-free grammar that generates all secondary
structures having a unique occurrence of one of
$\star \ca \ca$, $\oa \oa \star$, $\oa \star \ca$.  Since two
shift moves correspond to each of the expressions
$\star \ca \ca$, $\oa \oa \star$, $\oa \star \ca$, it follows
that the total number of $MS_2-MS_1$ (shift-only) moves, summed over all
structures for a homopolymer of length $n$ with $\theta=3$, is equal to
$2 \cdot [z^n] \widehat{S}(z)$.

As explained in \cite{Lorenz.jcb08} and \cite{flajoletBook}, it is possible 
to automatically transform the previous production rules into equations that
relate the corresponding generating functions, where we denote generating 
functions of $\widehat{S}(z)$, $\widehat{T}(z)$, $S(z)$, $R(z)$ by the same 
symbols used for the corresponding nonterminals $\widehat{S}$, $\widehat{T}$, 
$S$, $R$. This technique is known as DSV (Dyck, Sch\"utzenberger, Viennot) 
methodology in
\cite{Lorenz.jcb08}, or as the {\em symbolic method} in \cite{flajoletBook}
-- see Table~\ref{table:DSV}.  In this fashion, we obtain the following:
\begin{align*}
\widehat{S} &= z \widehat{S} + z^2 \widehat{S} +
z^2 S \widehat{S} + z^2 R \widehat{S} + \widehat{T} \\
\widehat{T} &= 2 z^3 R + 4 z^3 R S + 2 z^3 R S^2  +  z^3 R^2 +
z^3 S R^2\\
S &= z + z S + z^2 R + z^2 R S \\
R &= \theta + z R  + z^2 R  + z^2 R S \\
\theta &= z^3
\end{align*}
and by eliminating all variables except $\widehat{S}$ and $z$, we use
Mathematica to obtain the quadratic equation in $\widehat{S}$
having two solutions, for which the only solution analytic at $0$ is 
the following:
\begin{align}
\label{eqn:formalValueWidehatSforMS2theta3}
\widehat{S}(z) &= \widehat{S} = \frac{A+B\sqrt{P}}{C}
\end{align}
where
\begin{align*}
P &= 1 - 2 z - z^2 + z^4 + 3 z^6 + 2 z^7 + z^8 \\
A &= 3 - 15 z + 23 z^2 - 9 z^3 - z^4 - 9 z^5 + \\
& 23 z^6 - 25 z^7 + 7 z^8 - z^9 + 6 z^{10} - \\
& 8 z^{11} + 2 z^{12} + 2 z^{13} + 2 z^{14}  \\
B &= -3 + 12 z - 14 z^2 + 4 z^3 + 5 z^5 - 10 z^6 + \\
& 8 z^7 - 2 z^{10} \\
C &= 2 (-z^3 + 3 z^4 - z^5 - z^6 - z^7 + z^8 - \\
& 3 z^9 + z^{10} + z^{11} + z^{12})
\end{align*}
The {\em dominant singularity} $\rho$ of  $\widehat{S}(z)$ in
equation~(\ref{eqn:formalValueWidehatSforMS2theta3}) is the complex number
having smallest absolute value (or modulus) at which $\widehat{S}(z)$ is
not differentiable. For the functions in this paper, the dominant singularity
will always be the (complex) root of polynomial $P$ under the radical, having
smallest modulus -- since the square root function is not differentiable over
the complex numbers at zero.

Letting $\widehat{F}(z) = \frac{B \sqrt{P}}{C}$ and noting that the
dominant singularity $\rho = 0.436911$, a calculation shows that
\begin{align*}
\lim\limits_{z\rightarrow \rho} \widehat{F}(z) &=
\lim\limits_{z\rightarrow \rho} 
\frac{B \cdot \sqrt{P'} \cdot (1-z/\rho)^{1/2}}{C' \cdot (1-z/\rho)}\\
P' &= \frac{P}{1-z/\rho} \\
&=  1 + 0.288795 z - 0.339007 z^2 - \\
&  0.775919 z^3 - 0.775919 z^4 - 1.775919 z^5 - \\
& 1.064714 z^6 - 0.436911 z^7 \\
C' &= \frac{C}{1-z/\rho} \\
 &= -2 z^3 + 1.422410 z^4 + 1.255605 z^5 + \\
& 0.873822 z^6 + 2 z^8 - 1.422410 z^9 - \\
& 1.255605 z^{10} - 0.873822 z^{11}
\end{align*}
and so
\begin{align*}
\lim\limits_{z\rightarrow \rho} \widehat{F}(z) &= 0.684877 \cdot 
\lim\limits_{z\rightarrow \rho} \left( 1-z/\rho \right)^{-1/2} \\
&= 0.684877 \cdot 
\lim\limits_{z\rightarrow \rho} \left( 1-z/0.436911 \right)^{-1/2}
\end{align*}
Taking $\alpha=-1/2$ in the Flajolet-Odlyzko Theorem \cite{FlaOdl90},
we obtain:
\begin{align*}
[z^n] \widehat{F}(z) &\sim
\frac{0.684877}{\Gamma(1/2)}
\cdot n^{-1/2} \cdot \left( \frac{1}{\rho} \right)^n \\
&= 0.3864 \cdot n^{-1/2} \cdot 2.28879^n
\end{align*}
By Theorem~\ref{thm:numSecStrMS1degree}
the asymptotic number of secondary structures for
a homopolymer when $\theta=3$ is $0.713121 \cdot n^{-3/2} \cdot 2.28879^n$,
and noting that $\lim_{z \rightarrow \rho} A/C = -4.36723$, 
we have the following result.
\begin{theorem}
\label{thm:MS2minusMS1degree}
The asymptotic $MS_2-MS_1$ degree of $\mathcal{S}_n$ is
\begin{align*}
\frac{2[z^n]\widehat{\widehat{S}}(z)}{[z^n]S(z)} &\sim
\frac{2[z^n](A/C + B\sqrt{P}/C)(z)}{[z^n]S(z)} \sim
\frac{2[z^n]\widehat{F}(z)}{[z^n]S(z)} \sim
\frac{0.772801 \cdot n^{-1/2} \cdot 2.28879^n}
{0.713121 \cdot n^{-3/2} \cdot 2.28879^n} \\
&= 1.083688 \cdot n
\end{align*}
\end{theorem}
Adding the asymptotic values from 
Theorem~\ref{thm:numSecStrMS1degree} and 
Theorem~\ref{thm:MS2minusMS1degree}, we determine the $MS_2$ degree.
\begin{corollary}
\label{cor:degreeMS2t3}
The asymptotic $MS_2$ degree for the network $\mathcal{S}_n$ of
RNA structures is $1.557164 \cdot n$.
\end{corollary}
Using a Taylor series expansion at zero for the functions used to determine
both the $MS_1$ and $MS_2-MS_1$ degree, we have verified that the numerical
results for $\mathcal{S}_n$ are identical with those independently computed
by the dynamic programming C-implementations described in 
\cite{cloteJCC2015} and \cite{Clote.po15}.
We also note that the current approach is {\em much}
simpler than the program in \cite{Clote.po15}, although the latter is
more general, since it computes the $MS_2$ degree for any user-specified
RNA sequence. Using well-known methods, these asymptotic results can be
extended from homopolymers to RNA sequences with Watson-Crick and wobble
base pairs by using a ``stickiness model'', which stipulates the probability
$p$ that any two positions can form a base pair is defined by
\begin{align}
\label{eqn:stickiness}
p &= 2\left( p_A\,p_U + p_G\,p_U + p_G\,p_C \right)
\end{align}
where $p_A, p_C, p_G, p_U$ are user-specified nucleotide relative frequencies.
Since we consider shifts, we need an additional stickiness
parameter $q$, which specifies the probability that a shift
can occur between three randomly selected positions in which one
position is fixed, defined by
\begin{align}
\label{eqn:shiftstickiness}
q &= p_A\, p_U^2 + p_C\, p_G^2 + p_G \left( p_C^2 + p_U^2 + 2 p_C\,p_U\right) +
p_U\left( p_A^2+p_G^2+2p_A\,p_G \right)
\end{align}
By including stickiness parameters into our computations, we obtained values
presented in Table~\ref{table:MS1MS2valuesForThetaPQ}, which shows
asymptotic $MS_1$ and $MS_2$ degrees for a number of classes of RNA.

\section{Asymptotic $MS_2$ clustering coefficient}
\label{section:clusteringCoefficient}

Section~\ref{section:triangles} describes a grammar to count the
number of triangles for $\mathcal{S}_n$ with respect to
$MS_2$ moves, while Section~\ref{section:triples} 
describes a grammar to count two particular triples. 
Figure~S1 provides a schematic overview of the
2 types of possible triangles and 4 types of triples. To each type of
triangle or triple, there corresponds a {\em motif}, depicted as an
undirected graph on 3, 4 or 5 nodes.

\subsection{Counting triangles} 
\label{section:triangles}

We now define a nonambiguous context-free grammar $G$, that
generates all secondary structures containing a unique
triangle motif, where type A [resp. B] triangles are enumerated in 
Figure~S2 [resp. S3],
corresponding to triangle rules 1-3 [resp. 4-8] below.
Grammar $G$ has the terminal symbols 
$\op,\bullet, \cp,\langle, \star, \rangle$,
nonterminal symbols 
$S^{\bigtriangleup}, S_1,\ldots,S_8, S, R, X,\theta$, start symbol
$S^{\bigtriangleup}$ and the following production rules:
\begin{align*}
S^{\bigtriangleup} &\rightarrow S_1 \,|\, S_2 \,|\, S_3  \,|\,
S_4 \,|\, S_5 \,|\, S_6 \,|\, S_7 \,|\, S_8 \\
S &\rightarrow \bullet \,|\, S \bullet \,|\, \op R \cp  \,|\, S \op R \cp \\
R &\rightarrow \theta  \,|\, R \bullet \,|\, \op R \cp  \,|\, S \op R \cp \\
X &\rightarrow \lambda \,|\, S \\   
\theta &\rightarrow \bullet \bullet \bullet
\end{align*}
where $\lambda$ denotes the empty word, and
$S_1,\ldots,S_8$ are specified in the following 8 exhaustive 
and mutually exclusive cases. Note that
$S_1,\ldots,S_3$ generate structures containing type A triangles, while
$S_4,\ldots,S_8$ generate structures containing type B triangles.
After each rule explanation, we give both the grammar and the corresponding
DSV equations.

\subparagraph*{Rule 1 $\oa \star \ca$}
The following productions
generate all secondary structures $s$, such that
for $x<y<z$, it is the case that
$s \cup \{(x,y)\}$ and $s \cup \{(y,z)\}$ are also secondary structures,
hence form a triangle:
\begin{align*}
S_1 &\rightarrow
S_1  \bullet \,|\, \op S_1 \cp \,|\,
S \op S_1 \cp \,|\, S_1 \op R \cp  \,|\, 
X \oa R \star R \ca  \\
S_1 &= z S_1 + z^2 S_1 + z^2 S S_1 + z^2 R S_1 + X z^3 R^2 
\end{align*}

\subparagraph*{Rule 2 $\star  \ca \ca$}
The following productions generate all secondary structures $s$, such that
for $x<y<z$, it is the case that
$s \cup \{(x,y)\}$ and $s \cup \{(x,z)\}$ are also secondary structures,
hence form a triangle:
\begin{align*}
S_2 &\rightarrow
S_2  \bullet \,|\, \op S_2 \cp \,|\,
S \op S_2 \cp \,|\, S_2 \op R \cp  \,|\, X \star R \ca X \ca  \\
S_2 &= z S_2 +  z^2 S_2 +  z^2 S S_2 +  z^2 R S_2 +  X^2 z^3 R 
\end{align*}

\subparagraph*{Rule 3 $\oa \oa \star$}
The following productions generate all secondary structures $s$, such that
for $x<y<z$, it is the case that
$s \cup \{(x,z)\}$ and $s \cup \{(y,z)\}$ are also secondary structures,
hence form a triangle:
\begin{align*}
S_3 &\rightarrow
S_3  \bullet \,|\, \op S_3 \cp \,|\,
S \op S_3 \cp \,|\, S_3 \op R \cp  \,|\, X \oa X \oa R \star  \\
S_3 &= z S_3 +  z^2 S_3 +  z^2 S S_3 +  z^2 R S_3 +  X^2 z^3 R
\end{align*}

\subparagraph*{Rule 4 $\star \ca \ca \ca$}
The following productions generate all secondary structures $s$, such that
for $x<y<z<w$, it is the case that
$s \cup \{(x,y)\}$,
$s \cup \{(x,z)\}$ and
$s \cup \{(x,w)\}$ are also secondary structures,
hence the latter form a triangle:
\begin{align*}
S_4 &\rightarrow
S_4  \bullet \,|\, \op S_4 \cp \,|\,
S \op S_4 \cp \,|\, S_4 \op R \cp  \,|\, X \star R \ca X \ca X \ca  \\
S_4 &= z S_4 +  z^2 S_4 +  z^2 S S_4 +  z^2 R S_4 +  X^3 R z^4
\end{align*}

\subparagraph*{Rule 5 $\oa \oa \oa \star$}
For $x<y<z<w$, let $s_1=(x,w)$, $s_2= (y,w)$, $s_3=(z,w)$. 
The following productions generate all secondary structures $s$, such that
for $x<y<z$, it is the case that
$s \cup \{(x,w)\}$,
$s \cup \{(y,w)\}$ and
$s \cup \{(z,w)\}$ are also secondary structures,
hence the latter form a triangle:
\begin{align*}
S_5 &\rightarrow
S_5  \bullet \,|\, \op S_5 \cp \,|\,
S \op S_5 \cp \,|\, S_5 \op R \cp  \,|\, X \oa X \oa X \oa R  \star  \\
S_5 &= z S_5 +  z^2 S_5 +  z^2 S S_5 +  z^2 R S_5 +  X^3 z^4 R 
\end{align*}

\subparagraph*{Rule 6 $\oa \star \ca \ca$}
For $x<y<z<w$, 
the following productions generate all secondary structures $s$, such that
for $x<y<z$, it is the case that
$s \cup \{(x,y)\}$,
$s \cup \{(y,z)\}$ and
$s \cup \{(y,w)\}$ are also secondary structures,
hence the latter form a triangle:
\begin{align*}
S_6 &\rightarrow
S_6  \bullet \,|\, \op S_6 \cp \,|\,
S \op S_6 \cp \,|\, S_6 \op R \cp  \,|\, 
X \oa X \oa R \star R \ca  \\
S_6 &= z S_6 +  z^2 S_6 +  z^2 S S_6 +  z^2 R S_6 +  X^2 z^4 R^2 
\end{align*}

\subparagraph*{Rule 7 $\oa \oa \star \ca$}
For $x<y<z<w$, 
the following productions generate all secondary structures $s$, such that
for $x<y<z$, it is the case that
$s \cup \{(x,z)\}$,
$s \cup \{(y,z)\}$ and
$s \cup \{(z,w)\}$ are also secondary structures,
hence the latter form a triangle:
\begin{align*}
S_7 &\rightarrow
S_7  \bullet \,|\, \op S_7 \cp \,|\,
S \op S_7 \cp \,|\, S_7 \op R \cp  \,|\, 
 X \oa R \star R \ca X \ca  \\
S_7 &= z S_7 +  z^2 S_7 +  z^2 S S_7 +  z^2 R S_7 +  X^2 z^4 R^2 
\end{align*}

\subparagraph*{Rule 8 $\oa \star \ca$ bis}
The following productions generate all secondary structures $s$, such that
for $x<y<z$, it is the case that
$s \cup \{(x,z)\}$,
$s \cup \{(x,y)\}$ and
$s \cup \{(y,z)\}$ are also secondary structures,
hence the latter form a triangle.  This grammar is
identical to that in rule 1 above, with the exception that
$S_1$ is replaced by $S_8$. 

Let $S^{\bigtriangleup}(z)$ denote the generating function for the number of
structures containing a unique triangle motif, where 
$triA(z)$ [resp. $triB(z)$] is the generating function for the collection 
of structures containing a unique occurrence of type A [type B]
triangle, as treated in rules 1-3 [resp. rules 4-8].
We obtain the following compact form for the DSV equations for
the grammar $G$ that generates all structures containing a triangle:
\begin{align*}
S^{\bigtriangleup} &= triA + triB\\
triA &= triA \cdot z + X \cdot z \cdot triA \cdot z +  \\
& triA \cdot z \cdot R \cdot z + X \cdot z \cdot R \cdot z \cdot R \cdot z +  \\
& X \cdot z \cdot R \cdot z \cdot X \cdot z +
X \cdot z \cdot X \cdot z \cdot R \cdot z \\
triB &= triB \cdot z + X \cdot z \cdot triB \cdot z + \\ 
& triB \cdot z \cdot R \cdot z + X^3 z^4 R + X^3 z^4 R + \\
& X^2 z^4 R^2 + X^2 z^4 R^2 + X z^3 R^2 
\end{align*}
Using Mathematica, we determine the following.
\begin{align*}
[z^n]S^{\bigtriangleup}(z) &= 
0.870311 \cdot 2.28879^n \cdot n^{-1/2}
\end{align*}
By Theorem~\ref{thm:numSecStrMS1degree}, the asymptotic number of 
secondary structures is $0.713121 \cdot n^{-3/2} \cdot 2.28879^n$,
and so we have the following result.
\begin{theorem}
\label{thm:numberOfTriangles}
The asymptotic average number of triangles per structure is
\begin{align*}
\frac{[z^n]S^{\bigtriangleup}(z)}{[z^n]S(z)} &\sim 
\frac{0.870331 \cdot n^{-1/2} \cdot 2.28879^n}
{0.713121 \cdot n^{-3/2} \cdot 2.28879^n} \\
&\sim 1.220453 \cdot n
\end{align*}
\end{theorem}

\subsection{Counting triples}
\label{section:triples}

In this section, we describe a grammar for two particular triples.
Let $G$ be the grammar having terminal symbols
$\bullet, \op, \cp, \ob, \cb$, nonterminal symbols
$S^{\ddag}$, $S^{\dag}$, $S$, $R$, $X$, $\theta$, start symbol
$S^{\ddag}$, and productions given in equation~(\ref{eqn:grammarTriples})
below together with the following:
\begin{align*}
S &\rightarrow \bullet  \,|\, S \bullet \,|\, X \op R \cp  \nonumber \\
R &\rightarrow \theta \,|\, R \bullet \,|\, X \op R \cp \nonumber \\
X &\rightarrow \lambda \,|\, S \nonumber \\
\theta &\rightarrow \bullet \bullet \bullet
\end{align*}

\subsubsection*{Triple with motif $\ob \cb \ob \cb$ \mbox{ or } $\ob \ob \cb \cb$}
The following grammar generates all secondary structures $s$ that have
two special base pairs $(i,j)$ and $(x,y)$, designated by
$\ob \cb$, which are either sequential or nested, as shown in the first
two panels of Figure~S4.
For each structure $s$, which contains a unique occurrence of the
sequential motif $\ob \cb \ob \cb$  or of the nested motif
$\ob \ob \cb \cb$, we must count four possible triples:
(1) $\{ s_1,s_2,s_3 \}$, where $s_1=s - \{(i,j),(x,y)\}$, 
$s_2=s - \{(i,j)\}$, $s_3=s - \{(x,y) \}$.  
(2) $\{ s_1,s_2,s_3\}$, where $s_1=s$, $s_2=s - \{(i,j)\}$, 
$s_3=s - \{(x,y) \}$.  
(3) $\{ s_1,s_2,s_3\}$, where $s_1=s - \{(i,j)\}$, $s_2=s - \{ (i,j),(x,y) \}$,
$s_3=s$.
(4) $\{ s_1,s_2,s_3\}$, where $s_1=s - \{(x,y)\}$, $s_2=s - \{ (i,j),(x,y) \}$,
$s_3=s$.
For this reason, we multiply by $4$ the asymptotic number of structures
generated by the following grammar $G$, whose terminal symbols are
$\bullet, \op, \cp, \ob, \cb$, and nonterminal symbols are
$S^{\ddag}$, $S^{\dag}$, $S$, $R$, $X$, $\theta$ with 
start symbol $S^{\ddag}$. Here
$S^{\dag}$ [resp. $S^{\ddag}$] generates all structures with $1$
[resp. $2$] distinguished base pairs ($\ob$, $\cb$), so
$S^{\ddag}$ generates all structures containing a unique triple of
the form $\ob \cb \ob \cb$ or$\ob \ob \cb \cb$.
\begin{align}
\nonumber
S^{\ddag} &\rightarrow S^{\ddag} \bullet  \,|\, 
\op S^{\ddag}\cp  \,|\, S \op S^{\ddag} \cp  \,|\, S^{\ddag} \op R \cp  \,|\, \\
\nonumber
& \ob S^{\dag} \cb \,|\, S \ob S^{\dag} \cb \,|\, S^{\dag} \ob R \cb \,|\,  
S^{\dag} \op S^{\dag} \cp \\
\nonumber
S^{\dag} &\rightarrow S^{\dag} \bullet  \,|\, \op S^{\dag}\cp  \,|\, 
S \op S^{\dag} \cp  \,|\, S^{\dag} \op R \cp  \,|\, \\
\label{eqn:grammarTriples}
& \ob R \cb \,|\, S \ob R \cb  
\end{align}
When applying the Flajolet-Odlyzko Theorem in the current case, we have
$\rho=0.436911$ and $\alpha=-3/2$.  A computation shows that
\begin{align*}
\lim_{z \rightarrow \rho} S^{\ddag}(z) &=
0.0177098 \left(1-z/\rho)\right)^{-3/2}\\
[z^n]S^{\ddag}(z) &\sim
0.0199834 \cdot n^{1/2} \cdot 2.28879^n \\
\frac{[z^n]S^{\ddag}(z)}{[z^n]S(z)}
 &\sim 
\frac{0.0199834 \cdot n^{1/2} \cdot 2.28879^n} 
{0.713121 \cdot n^{-3/2} \cdot 2.28879^n} \\
&\sim 0.0280225 \cdot n^2
\end{align*}
As mentioned, the number of triples contributed in the current case
is {\em 4 times} the last value.
Thus the expected number of triples involving a
structure containing $\ob \cb \ob \cb$ or $\ob \ob \cb \cb$ is
$4 \cdot 0.0280224 \cdot n^2 = 0.1120896 \cdot n^2$.
\begin{theorem}
\label{thm:typeAtriples1and2}
The asymptotic average number of triples per structure, for the triples
described in this section, is
\begin{align*}
\frac{4[z^n]S^{\ddag}(z)} {[z^n]S(z)} 
&\sim 0.11209 \cdot n^2
\end{align*}
\end{theorem}
From Theorems \ref{thm:numberOfTriangles} and 
\ref{thm:typeAtriples1and2}, we obtain an upper bound for the global clustering
coefficient, defined in equation~(\ref{eqn:globalClusteringCoefficient}).
\begin{theorem}[Bound on global clustering coefficient]
\label{thm:upperboundClusteringCoeff}
\begin{align*}
\mathfrak{C}_{g}(G)
&= \frac{3 \times \mbox{number of triangles}}
    {\mbox{number of connected triples}} 
= O \left( \frac{1}{n} \right)
\end{align*}
and hence the family $\mathcal{S}_n$, $n=1,2,3,\ldots$ of RNA secondary
structures is not small-world.
\end{theorem}

\section{Discussion}
\label{section:discussion}

In this paper, we have used methods from algebraic combinatorics
\cite{flajoletBook} to determine the asymptotic average degree
and asymptotic clustering coefficient of the $MS_2$ network $\mathcal{S}_n$
of RNA secondary structures.  Since the clustering
coefficient is not bounded away from zero, it follows that the family
$\mathcal{S}_n$, $n=1,2,3,\ldots$, of networks is not small-world.
In contrast, \cite{Wuchty.nar03} had shown by exhaustive enumeration of
low energy RNA secondary structures of {\em E. coli} phe-tRNA, that
the RNA network displays {\em small-world} properties. The seeming
contradiction of our rigorous result is due to the fact that the notion
of {\em finite} small-world network is not precisely defined due to 
absence of an exact bound for average path length and for
clustering coefficient.  

\begin{figure*}
\centering
\includegraphics[width=\textwidth]{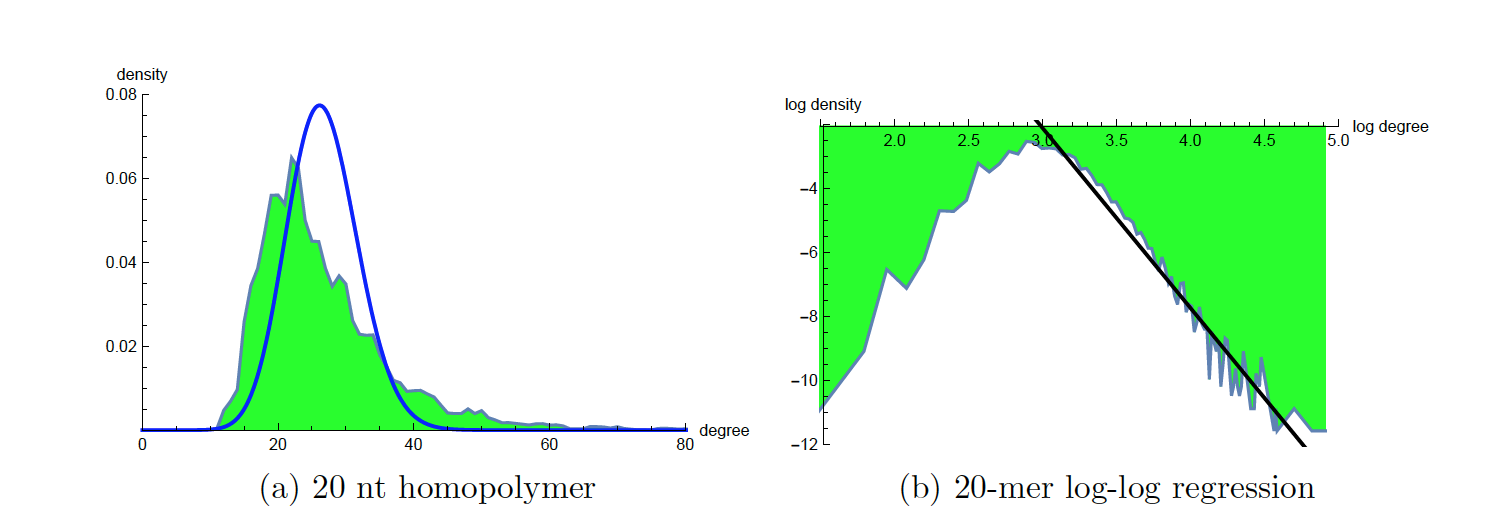}
\caption{(a) $MS_2$-degree distribution for the 106,633 secondary 
structures for a 
20-nt homopolymer with $\theta=3$ (green shaded curve), with Poisson
distribution of the same mean.
(b) Plot of $\ln(\mbox{density})$ as a function
of $\ln(\mbox{degree})$ for the degree distribution for $MS_2$
connectivity of the
20-nt homopolymer with $\theta=3$, for degrees $4,\ldots,136$.
The distribution tail appears to satisfy
a power-law with exponent $\approx -5.6247$, i.e.
$p(x) \propto  x^{-5.6247}$,
where $x$ is degree and $p(x)$ is the relative frequency of the number of
nodes having degree $x$ (regression equation for log-log plot is
$\ln(p(x)) = 14.7589 - 5.6247 \cdot x$).
}
\label{fig:degreeDist20mer}
\end{figure*}

A related question is whether RNA secondary structure networks are
{\em scale-free}, i.e. whether the tail of the degree density follows a
power-law distribution.
Figure~\ref{fig:degreeDist20mer} shows the $MS_2$ degree density for a 20-nt
homopolymer, computed by brute-force enumeration, together with 
a least-squares linear fit for the tail of the log-log plot, which suggests that
the distribution tail is given by a power-law with exponent $\approx -5.6247$.
A forthcoming paper by the last author settles this question
(see the {\tt arXiv} preprint 1807.00215).

\section*{Acknowledgments}
This research was supported in part by National Science Foundation grant
DBI-1262439 to PC and the
French/Austrian RNALands project (ANR-14-CE34-0011 and FWF-I-1804-N28)
to YP.  Any opinions, findings,
and conclusions or recommendations expressed in this material are
those of the authors and do not necessarily reflect the views of the
National Science Foundation.

\bibliographystyle{plain}

\hfill \clearpage

\begin{table}
\begin{tabular}{|l|l|}
        \hline
        \mbox{Type of nonterminal} & \mbox{Generating function} \\ \hline
        $A \to B \; | \; C$ & $A(z) = B(z) + C(z)$\\
                $A \to B\,C$        & $A(z) = B(z)C(z)$\\
                $A \to t$           & $A(z) = z$\\
                $A \to \lambda$     & $A(z) = 1$ \\ \hline
\end{tabular}
\caption{Translation between context-free grammars and generating functions.
Here, $G=(V,\Sigma,S,R)$ is a given context-free grammar,
$A,B,C$ are any nonterminal symbols in $V$, 
$t$ is a terminal symbol in $\Sigma$, and $\lambda$ denotes the empty
word. The generating functions
for the languages $L(A)$, $L(B)$, $L(C)$ are respectively denoted by
$A(z)$, $B(z)$, $C(z)$.}
\label{table:DSV}
\end{table}

\begin{table}
\begin{tabular}{@{}c c c c c c c@{}}
\toprule
Move set& $\theta$ & {\tt hp} & {\tt wc} & {\tt wcw} & {\tt wc tRNA} & {\tt wcw tRNA}\\
\midrule
$\MSOne$& $1$& 0.55279 & 0.42265 & 0.46158 & 0.42157 & 0.46021 \\
$\MSOne$&$3$& 0.47348 & 0.35130 & 0.38531 & 0.35038 & 0.38408 \\
$\MSTwo-\MSOne$&$1$& 1.44721 & 0.57735 & 0.84796 & 0.57985 & 0.85119 \\
$\MSTwo-\MSOne$&$3$& 1.08369 & 0.42908 & 0.31409 & 0.21550 & 0.63055 \\
$\MSTwo$&$1$ & 2.00000 &1.00000 &1.30954 &1.00142 &1.31140 \\
$\MSTwo$&$3$ & 1.55717 &0.43527 &0.32336 &0.22162 &0.63971 \\
\bottomrule
\end{tabular}
\caption{The asymptotic expected degree of the network 
$\mathcal{S}^{\theta,p}_n$ of secondary structures for move sets
$\MSOne$ and $\MSTwo$ for different values of threshold $\theta$ and
base pair and triple stickiness parameters $p$ and $q$, defined in
Equations~\eqref{eqn:stickiness} and \eqref{eqn:shiftstickiness} respectively.
Five models are considered: {\tt hp} -- homopolymer model  with $p=q=1$;
{\tt wc} -- Watson-Crick pairing model with uniform compositional
frequency $p_A=p_C=p_G=p_U=\frac{1}{4}$, hence $p= 
\frac{1}{4}$, and
$q=
 0.0625$);
{\tt wcw} -- Watson-Crick and wobble pairing model  with uniform
compositional frequency, hence 
$p= 
\frac{3}{8}$, and
$q = 
0.15625$;
{\tt wc tRNA} -- Watson-Crick base pairing model based on the compositional 
frequencies ($p_A = \frac{1288}{4534}$,
$p_U = \frac{1029}{4534}$, $p_G = \frac{1223}{4534}$, $p_U = \frac{994}{4534}$)
observed in family RF00005 of 4,534 tRNAs
in the Rfam 12.0 database~\cite{Nawrocki.nar14}, hence $p=0.259427$, $q=0.066394$;
{\tt wcw tRNA} -- Watson-Crick and Wobble base pairing model using 
compositional frequency of RF00005, so that 
$p = 0.377699$, $q = 0.158476$.
}
\label{table:MS1MS2valuesForThetaPQ}
\end{table}
\hfill \clearpage

\hfill \clearpage

\appendix
\renewcommand\thefigure{S\arabic{figure}}
\setcounter{figure}{0}
\thispagestyle{empty}
\setcounter{page}{1}
\input{appendix}

\hfill \clearpage

\input{suppFigures}

\end{document}

%% file: macros.tex
\usepackage{graphicx,color,wrapfig}
\usepackage{alphalph}
\usepackage{amssymb}
\usepackage{amsmath}
\usepackage{url}
\usepackage{xspace}
\usepackage{graphics}
\usepackage{caption}
\usepackage{booktabs} 
\usepackage{subcaption}
\usepackage{relsize} 

\usepackage{verbatim}
\usepackage{tkz-graph}
\usepackage{fullpage}
\usetikzlibrary{positioning}
\usepackage{floatrow}

\newcommand{\triangles}{{\Delta}}
\newcommand{\triA}{{\triangles_A}}
\newcommand{\triB}{{\triangles_B}}
\newcommand{\triples}{{\Lambda}}
\newcommand{\tripA}{{\triples_A}}
\newcommand{\tripB}{{\triples_B}}
\newcommand{\tripC}{{\triples_C}}
\newcommand{\tripCdisc}{{\triples_{C\,disc}}}
\newcommand{\tripCD}{{\triples_{CD}}}

\newcommand{\obb}{\,\mbox{\bf\texttt{$\{$}}\,}
\newcommand{\cbb}{\,\mbox{\bf\texttt{$\}$}}\,}
\newcommand{\ob}{\,\mbox{\bf\texttt{[}}\,}
\newcommand{\cb}{\,\mbox{\bf\texttt{]}}\,}
\newcommand{\op}{\,\mbox{\bf\texttt{(}}\,}
\newcommand{\cp}{\,\mbox{\bf\texttt{)}}\,}
\newcommand{\oa}{\,\mbox{\bf\texttt{$\langle$}}\,}
\newcommand{\ca}{\,\mbox{\bf\texttt{$\rangle$}}\,}

\usepackage{textcomp}  

\newtheorem{theorem}{Theorem}
\newtheorem{corollary}[theorem]{Corollary}

\chardef\other=12

\def\mmakeactive#1{\catcode`#1=\active\ignorespaces}

{
\mmakeactive\
\gdef\obeywhitespace{%
  \mmakeactive\^^M %
  \let^^M=\NewLine %
  \aftergroup\removebox %
  \obeyspaces %
}}

\def\NewLine{\par\indent}
\def\removebox{\setbox0=\lastbox}

\def\|{|}

%% file: appendix.tex
\section{Supplementary Information}

\subsection*{Complete grammar}
\label{subsection:completeGrammarClusteringCoefficient}

In this supplement, we list the complete grammar for all triangles and
connected triples, together with the corresponding functional equations.
We do not provide any detailed explanation for the complete listing of
all possible triples or the complete grammar, since our intent is to provide
suggestive supplementary figures and a general orientation for the
reader wishing to work through the Mathematica code, available upon
request, for our computation of the asymptotic clustering coefficient 
value $20.7728/n$.

\subsection{Triangles and type A triples}

Figure~\ref{fig:trianglesAndTriples} provides an overview of the
classification of triangle motifs (types A and B) and non-triangle
triple motifs (types A,B,C,D). In this section, we describe grammar
rules for triangles and type A triples -- see
Figure~\ref{fig:typeAtriangles} for type A triangles,
Figure~\ref{fig:typeBtriangles} for type B triangles, and
Figure~\ref{fig:typeAtriples} for type A triples. Triangles of type A and
B have been explained in the main text; 
type A triples are constituted by structures $s_1,s_2,s_3$,
where $s_2,s_3$ are obtained from $s_1$ by a base pair addition with the
property that there is a no $MS_2$ move between $s_2$ and $s_3$.

In the following production rules,
nonterminal $S$ generates all non-empty secondary structures; $R$ 
[resp. $S2$] generates all secondary structures of length at least $3$
[resp. $2$].  Nonterminal $PK$ generates all structures containing
a unique pseudoknot (crossing base pairs) of the form $\ob \obb \cb \cbb$,
where a second type of bracket is necessary to properly parse the expression.
Nonterminal $\triA$ [resp. $\triB$] generates all structures containing a 
unique occurrence of a type $A$ [resp. $B$] triangle, and nonterminal
$\tripA$ generates all structures containing a 
unique occurrence of a type $A$ triple.
Nonterminal $S^{\dag}$ [resp. $S^{\ddag}$]
generates all secondary structures
containing a unique occurrence of a motif that has exactly 
$1$ [resp. $2$] extended connected components. Thus $S^{\ddag}$ generates
all structures having a unique occurrence of motif 1 or 2 for type A triple.

\begin{align*}
S &\rightarrow \bullet \,|\, S \bullet \,|\, X \op R \cp \\
R &\rightarrow \theta \,|\, R \bullet \,|\, X \op R \cp \\
S2 &\rightarrow \bullet \bullet \,|\, R \\
\theta &\rightarrow \bullet  \bullet \bullet \\
X &\rightarrow \lambda \,|\, S  \\
 \triA  &\rightarrow  \triA  \bullet \,|\, X \op  \triA  \cp \,|\,  \triA  \op R \cp \,|\,
     X \oa R \bigstar R \ca \,|\, X \bigstar R \ca X \ca \,|\,
     X \oa X \oa R \bigstar \qquad  \mbox{(type A triangles)} \\
 \triB  &\rightarrow  \triB  \bullet \,|\, X \op  \triB  \cp \,|\, 
      \triB  \op R \cp \,|\,
     X \bigstar R \ca X \ca X \ca  \,|\, X \oa X \oa X \oa R  \bigstar \,|\, \\
&     X \oa X \oa R \bigstar R \ca \,|\, X \oa R \bigstar R \ca X \ca \,|\,
     X \bigstar R \bigstar R \bigstar \qquad \mbox{(type B triangles)} \\
\triangles &\rightarrow  \triA  \,|\,  \triB  \qquad  \mbox{(all triangles)}  \\
PK &\rightarrow  PK \bullet \,|\, X \op PK \cp \,|\, PK \op R \cp \,|\,
   X \ob \obb S2 \cb \cbb \,|\, X \ob S2 \obb \cb S2 \cbb \,|\, \\
&   X \ob \obb S2 \cb S \cbb \,|\, X \ob S \obb S2 \cb \cbb \,|\,
   X \ob S \obb S \cb S \cbb \\
\tripA &\rightarrow PK \,|\, S^{\ddag} \qquad  \mbox{(Type A triples)} \\ 
S^{\dag} &\rightarrow S^{\dag}   \bullet \,|\, X \op S^{\dag} \cp \,|\, 
          S^{\dag} \op R \cp \,|\,  X \ob R \cb \\
S^{\ddag} &\rightarrow S^{\ddag}   \bullet \,|\, X \op S^{\ddag} \cp \,|\, 
          S^{\ddag} \op R \cp \,|\,  X \ob S^{\dag} \cb \,|\, S^{\dag} \ob R \cb \,|\,
          S^{\dag} \op S^{\dag} \cp \qquad \mbox{(Note: $\times 4$)} \\
\end{align*}

\subsection{Type B triples}

Figure~\ref{fig:typeBtriples} depicts type B triples, defined to be of 
the form $s_1,s_2,s_3$, where $s_1,s_2$ are connected
by a shift, as are $s_1,s_3$, but $s_2,s_3$ are not connected by any
$MS_2$ move.
Nonterminal $\tripB$ generates all secondary structures containing
a unique occurrence of one of the 12 type B triple motifs in
Figure~\ref{fig:typeBtriples}. Nonterminals $R_1,\ldots,R_{12}$
generate respectively the collection of secondary structures containing
a unique occurrence of motif $1,\ldots,12$. Note that $S2$ is not the same
as $R_2$ -- the former nonterminal $S2$ simply generates all secondary 
structures of length $2$ or greater.

\begin{align*}
\tripB &\rightarrow \tripB \bullet \,|\, X \op \tripB \cp \,|\, \tripB \op R \cp \,|\,
 R_1 \,|\, \cdots \,|\, R_{12} \qquad \mbox{(type B triples)}\\
R_1 &\rightarrow X \bigstar R \bigstar R \ca X \ca \\
R_2 &\rightarrow X \bigstar X \oa R \bigstar X \ca \\
R_3 &\rightarrow X \bigstar R \bigstar S2 \ca \ca \,|\, X \bigstar R \bigstar \ca S2 
\,|\, X \bigstar R \bigstar S \ca S \ca \\ 
R_4 &\rightarrow X \bigstar \oa S2 \ca \bigstar \,|\, X \bigstar S2 \oa \ca S2 \bigstar \,|\, 
 X \bigstar \oa S2 \ca S \bigstar \,|\, X \bigstar S \oa S2 \ca \bigstar \,|\, 
 X \bigstar S \oa S \ca S \bigstar  \\
R_5 &\rightarrow X \oa X \bigstar R \bigstar X \ca \\
R_6 &\rightarrow X \oa X \bigstar R \ca X \bigstar \\
R_7 &\rightarrow X \bigstar R \ca X \bigstar R \ca  \\
R_8 &\rightarrow X \bigstar R \ca X \oa  R \bigstar  \\
R_9 &\rightarrow X \oa R \bigstar R \bigstar R \ca   \\
R_{10} &\rightarrow X \oa X \oa R \bigstar R \bigstar   \\
R_{11} &\rightarrow X \oa R \bigstar X \oa R \bigstar   \\
R_{12} &\rightarrow X \oa S2 \oa \bigstar R \bigstar \,|\,
 X \oa S \oa S  \bigstar R \bigstar \,|\,
 X \oa \oa S2  \bigstar R \bigstar \\
\end{align*}

\subsection{Type C and D triples}

Figure~\ref{fig:typeCDtriples} depicts triples of type C and D, defined
to be of the form
are of the form $s_1,s_2,s_3$, where $s_1,s_2$ are connected
by a shift, and $s_3$ is obtained from $s_1$ by a base pair addition,
but $s_2,s_3$ are not connected by any $MS_2$ move.

Recall that an undirected graph is connected if there is a path between
any two nodes, and that any undirected graph is either connected or can
be decomposed into connected components.
Note that the motif for each type B triple is connected, and
that there are both connected and unconnected motifs
for triples of types C and D, but that each unconnected motif has exactly
two connected components.
A motif is defined to contain a pseudoknot if its arc cross -- this
does not occur for any triangle, but it does happen
in the type A triple (3), type B triples (3,4,5,12), and
type C,D triples (2,4,9,10,12,13,14,15,22,23).

If there exists an edge $(x,y)$ belonging to one connected
component of a motif, and
another edge $(u,v)$ belonging to a different connected component of
the same motif, and if $(x,y),(u,v)$ creates a pseudoknot, then both
components are merged into a single {\em extended connected component}.
With this definition,  there are two type A motifs that have one
extended connected component (3,4), and two type A motifs that have
two extended connected component (1,2); all type B motifs are connected;
there are 10 type C,D motifs having one extended connected component
(2, 4, 9, 10, 12, 13, 14, 15, 22, 23), and 15 type C,D motifs having
two extended connected components
(1, 3, 5, 6, 7, 8, 11, 16, 17, 18, 19, 20, 21, 24, 25).
Computations using DSV equations deriving from the grammars show that
if a motif has a single extended connected component, then
the asymptotic average number of triples per structure is  $O(n)$, while
if it has two extended connected components, then
the asymptotic average number is $O(n^2)$. Thus, the
asymptotic average number of triangles per structure is $O(n)$;
the asymptotic average number of triples per structure for type
A triples (1,2) is $O(n^2)$ while that of type A triples (3,4) is
$O(n)$.  Without further computations, it immediately follows that
the asymptotic clustering coefficient is $O\left(\frac{1}{n}\right)$.
Nonetheless, we note more generally, that
the asymptotic average number of triples per structure for any
motif having two extended connected components is $O(n^2)$, while that
for a motif having only one extended connected component is $O(n)$.
Finally, since the $O\left(\frac{1}{n}\right)$ asymptotic clustering
coefficient for the homopolymer implies an $O\left(\frac{1}{n}\right)$
asymptotic clustering coefficient for any nontrivial stickiness factors
$p,q$, we will not perform computations for these cases.

We now present the corresponding grammar, where
nonterminal $triples$ generates all secondary structures containing
a unique occurrence of one of the 25 type C or D triple motifs, as depicted in
Figure~\ref{fig:typeCDtriples}. 
Nonterminals $\tripC_1$, $\tripC_2$, $\tripC_{367}$, $\tripC_4$, $\tripC_5$, $
  \tripC_8$, $\tripC_9$, $\tripC_{10}$, $\tripC_{11}$, $\tripC_{12}$, $
  \tripC_{13}$, $\tripC_{14}$, $\tripC_{15}$, 
$\tripC_{\mbox{\tiny disc}}$, $\tripC_{20}$, $\tripC_{21}$, $\tripC_{22}$, $\tripC_{23}$
generate respectively the collection of secondary structures containing
a unique occurrence of motif $1,\ldots,25$, whereby
nonterminal $\tripC_{367}$ is the same rule for motif 3,6 and 7, and
nonterminal  $\tripC_{\mbox{\tiny disc}}$ is the same rule for all 
{\em disconnected \underline{successive}} motifs -- 
i.e. the 6 type C and D triple motifs
16, 17, 18, 19, 24, 25 which have exactly 2 extended connected components of
successive form $\ob \cb \ob \cb$. 

\begin{align*}
\tripC  &\rightarrow   \tripC \bullet \,|\, X \op \tripC \cp \,|\, \tripC \op R \cp \,|\,
  \tripC_1 \,|\, \tripC_2 \,|\, \tripC_{367} \,|\, \\
& \tripC_4 \,|\, \tripC_5 \,|\,
  \tripC_8 \,|\, \tripC_9 \,|\, \tripC_{10} \,|\, \tripC_{11} \,|\, \tripC_{12} \,|\, \\
&  \tripC_{13} \,|\, \tripC_{14} \,|\, \tripC_{15} \,|\,
  \tripC_{\mbox{\tiny disc}} \,|\, \tripC_{20} \,|\, \tripC_{21} \,|\, \tripC_{22} \,|\, \tripC_{23}\\
\tripC_1 &\rightarrow X \bigstar S^{\dag} \ca X \ca \qquad 
\mbox{(Note: $\times 4$)} \\
\tripC_2 &\rightarrow X \bigstar \oa S2 \ca \ca X \ca \,|\, 
  X \bigstar S2 \oa \ca S2 \ca X \ca \,|\,
  X \bigstar \oa S2 \ca S \ca X \ca \,|\, \\
& X \bigstar S \oa S2 \ca \ca X \ca \,|\, X \bigstar S \oa S \ca S \ca X \ca \\
\tripC_{367} &\rightarrow X \ob  \triA  \cb \qquad \mbox{(Note: $\times 4$)}\\
\tripC_4 &\rightarrow X \oa X \bigstar R \ca X \ca X \ca \\
\tripC_5 &\rightarrow X \bigstar R \ca S^{\dag} \ca \qquad 
\mbox{(Note: $\times 4$)} \\
\tripC_8 &\rightarrow   X \oa R \bigstar S^{\dag} \ca  \qquad 
\mbox{(Note: $\times 4$)} \\
\tripC_9 & \rightarrow X \oa \oa S2 \bigstar R \ca X \ca \,|\, 
X \oa S2 \oa \bigstar R \ca X \ca \,|\, X \oa S \oa S \bigstar R \ca X \ca \\
\tripC_{10} &\rightarrow X \oa X \oa X \oa R \bigstar X \ca \\
\tripC_{11} &\rightarrow X \oa X \oa S^{\dag} \bigstar \qquad 
\mbox{(Note: $\times 4$)} \\
\tripC_{12} &\rightarrow X \bigstar R \ca X \oa \ca S2 \ca \,|\, 
  X \bigstar R \ca X \oa S2 \ca \ca \,|\, X \bigstar R \ca X \oa S \ca S \ca \\
\tripC_{13} &\rightarrow  X \oa X \oa R \bigstar \ca S2 \ca \,|\, 
  X \oa X \oa R \bigstar S2 \ca \ca \,|\, X \oa X \oa R \bigstar S \ca S \ca \\
\tripC_{14} &\rightarrow  X \oa R \bigstar \oa S2 \ca \ca \,|\,
  X \oa R \bigstar S2 \oa \ca S2 \ca \,|\, \\
&  X \oa R \bigstar \oa S2 \ca S \ca \,|\,
  X \oa R \bigstar S \oa S2 \ca \ca \,|\, X \oa R \bigstar S \oa S \ca S \ca \\
\tripC_{15} &\rightarrow  X \oa X \oa \oa S2 \ca \bigstar \,|\,
  X \oa X \oa S2 \oa \ca S2 \bigstar  \,|\,
  X \oa X \oa \oa S2 \ca S \bigstar \,|\, \\
&  X \oa X \oa S \oa S2 \ca \bigstar \,|\,
  X \oa X \oa S \oa S \ca S \bigstar \\
\tripC_{\mbox{\tiny disc}} &\rightarrow   \triA  \ob R \cb \,|\,
   \triA  \op S^{\dag} \cp \,|\, S^{\dag} \oa R \bigstar R \ca \,|\,
  S^{\dag} \oa X \oa R \bigstar \,|\,
  S^{\dag} \bigstar R \ca X \ca \,|\, \\
&  S^{\dag} \op X \oa R \bigstar R \ca X \cp \,|\, 
  S^{\dag} \op X \bigstar R \ca X \ca X \cp \,|\,
  S^{\dag} \op X \oa X \oa R \bigstar X \cp \\
\tripC_{20} &\rightarrow  X \oa S^{\dag} \bigstar R \ca \qquad 
\mbox{(Note: $\times 4$)} \\
\tripC_{21} &\rightarrow  X \oa S^{\dag} \oa R \bigstar  \qquad 
\mbox{(Note: $\times 4$)}\\
\tripC_{22} &\rightarrow   X \oa \oa S2 \ca \bigstar R \ca \,|\,
  X \oa S2 \oa \ca S2 \bigstar R \ca \,|\,
  X \oa \oa S2 \ca S \bigstar R \ca \,|\, \\
&  X \oa S \oa S2 \ca \bigstar R \ca \,|\,
  X \oa S \oa S \ca S \bigstar R \ca \\
\tripC_{23} &\rightarrow  X \oa S2 \oa \ca X \oa R \bigstar \,|\, 
  X \oa \oa S2 \ca X \oa R \bigstar \,|\, X \oa S \oa S \ca X \oa R \bigstar \\
\end{align*}
Finally, the collection of all secondary structures having a unique
motif for a connected triple (both non-triangular triples of types
A,B,C,D or deriving from triangles) is generated by the rule
\begin{align*}
\triples &\rightarrow \tripA \,|\, \tripB \,|\, \tripC \,|\, triangles \qquad 
\mbox{(Note: triangles must be multiplied by $3$)}
\end{align*}

This gives rise to the following functional equations for {\em all} connected
triples, both non-triangular triples, as well as 3 triples associated
with each triangle. Since rules $R_1,\ldots,R_{12}$ are written as
$R1,\ldots,R12$, we write {\tt S2} in place of $R2$, which latter had been
defined by $R2 \rightarrow \bullet \bullet \,|\, R$.
\begin{align*}
S^{\ddag} &= S^{\ddag} \cdot z + X \cdot z \cdot S^{\ddag} \cdot z + S^{\ddag} \cdot z \cdot R \cdot z + X \cdot z \cdot S^{\dag} \cdot z + 
  S^{\dag} \cdot z \cdot R \cdot z + S^{\dag} \cdot z \cdot S^{\dag} \cdot z \\
 S^{\dag} &= 
 S^{\dag} \cdot z + X \cdot z \cdot S^{\dag} \cdot z + S^{\dag} \cdot z \cdot R \cdot z + X \cdot z \cdot R \cdot z\\
 S &= 
 z + S \cdot z + X \cdot z \cdot R \cdot z\\
 R &= \theta + R \cdot z + X \cdot z \cdot R \cdot z \\
 \theta &= z \cdot z \cdot z\\
 X &= 1 + S \\
 S2 &= z \cdot z + R\\
 PK &= 
 PK \cdot z + X \cdot z \cdot PK \cdot z + PK \cdot z \cdot R \cdot z + X \cdot z^4 (S2 + S2^2 + 2 S S2 + S^3)\\
 \tripA &= PK + 4 \cdot S^{\ddag} \\
 R1 &= X^2 z^4 R^2\\
 R2 &= X^3  z^4 R\\
 R3 &= X z^4 R (2 S2 + S^2)\\
 R4 &= X z^4 (S2 + S2^2 + 2 S S2 + S^3)\\
 R5 &= X^3 z^4 R\\
 R6 &= X^3 z^4 R\\
 R7 &= X^2 z^4 R^2\\
 R8 &= 
 X^2 z^4 R^2\\
 R9 &= X z^4 R^3\\
 R10 &= X^2 z^4 R^2\\
 R11 &= 
 X^2 z^4 R^2\\
 R12 &= X z^4 (2 S2 + S^2)\\
\end{align*}
together with the following, where we write $\tripC$ in place of $\tripCD$
for reasons of brevity
\begin{align*}
 \tripB &= 
 R1 + R2 + R3 + R4 + R5 + R6 + R7 + R8 + R9 + R10 + R11 + R12 +  \\
&  \tripB z + X z^2 \tripB + \tripB z^2 R\\
 \tripC_{1} &= X z S^{\dag} z X z\\
 \tripC_{2} &= 
 X z^5 (S2 + S2^2 + 2 S S2 + S^3)\\
 \tripC_{367} &= X z  \triA  z \\
 \tripC_{4} &= 
 X^4 z^5 R\\
 \tripC_{5} &= X z R z S^{\dag} z\\
 \tripC_{8} &= X z R z S^{\dag} z\\
 \tripC_{9} &= 
 X^2 z^5 (2 S2 + S^2)\\
 \tripC_{10} &= X^4 z^5 R\\
 \tripC_{11} &= 
 X^2 z^3 S^{\dag}\\
 \tripC_{12} &= X^2 z^5 R (2 S2 + S^2) \\
 \tripC_{13} &= 
 X^2 z^5 R (2 S2 + S^2) \\
 \tripC_{14} &= 
 X z^5 R (S2 + S2^2 + 2 S S2 + S^3)\\
 \tripC_{15} &= 
 X^2 z^5 (S2 + S2^2 + 2 S S2 + S^3)\\
 \tripCdisc &= 
  \triA  z R z + 
   \triA  z S^{\dag} z + (S^{\dag} z R z R z + S^{\dag} z X z R z + 
     S^{\dag} z R z X z) (1 + X^2 z^2) \\
 \tripC_{20} &= X z^3 R S^{\dag}\\
 \tripC_{21} &= 
 X z^3 R S^{\dag}\\
 \tripC_{22} &= X z^5 R (S2 + S2^2 + 2 S S2 + S^3)\\
 \tripC_{23} &= X^2 z^5 R (2 S2 + S^2)\\
\tripC &= \tripC \cdot z + X \cdot z \cdot \tripC \cdot z + 
  \tripC \cdot z \cdot R \cdot z + 4 \cdot \tripC_{1} + \\
& \tripC_{2} + 4 \cdot \tripC_{367} + \tripC_{4} + 4 \cdot \tripC_{5} + 
 4 \cdot \tripC_{8} + \tripC_{9} + \\
& \tripC_{10} + 4 \cdot \tripC_{11} + \tripC_{12} + \tripC_{13} + 
 \tripC_{14} + \tripC_{15} + \\
& \tripCdisc +  4 \cdot \tripC_{20} + 4 \cdot \tripC_{21} + \tripC_{22} + \tripC_{23}\\
\triA  &= 
  \triA  \cdot z + X \cdot z \cdot  \triA  \cdot z +  \triA  \cdot z \cdot R \cdot z + 
  X \cdot z \cdot R \cdot z \cdot R \cdot z  + \\
& X \cdot z \cdot R \cdot z \cdot X \cdot z + 
  X \cdot z \cdot X \cdot z \cdot R \cdot z\\
\triB  &= 
  \triB  \cdot z + X \cdot z \cdot  \triB  \cdot z + 
  \triB  \cdot z \cdot R \cdot z + X^3 z^4 R + X^3 z^4 R +  \\
&  X^2 z^4 R^2 + X^2 z^4 R^2 + X z^3 R^2\\
\triangles &= 
  \triA  +  \triB \\
\triples &= \tripA + 4 \cdot \tripB + \tripC + 3 \cdot \triangles 
\end{align*}

%% file: suppFigures.tex
\begin{figure}
\centering
\includegraphics[width=\textwidth]{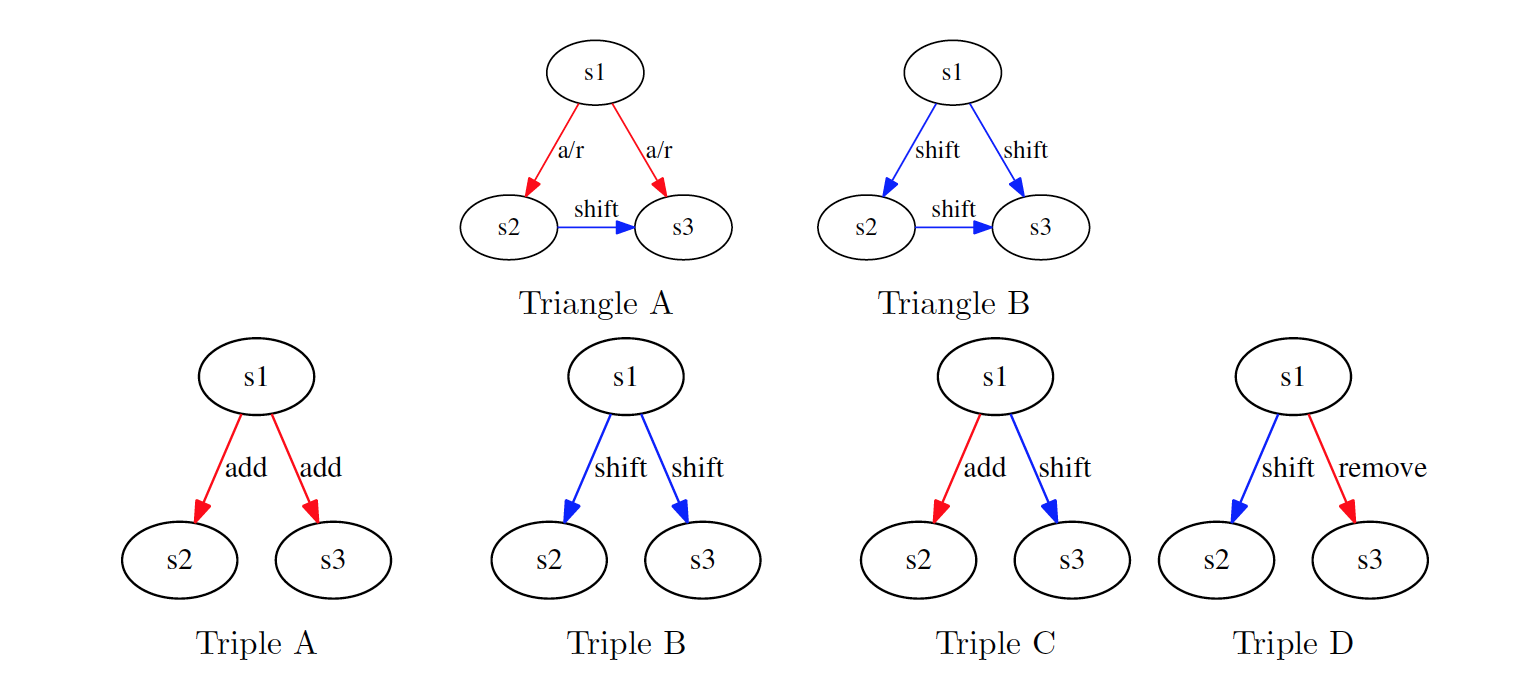}
\caption{Complete listing of all possible moves for 
triangles and non-triangular 
connected triples with a designated first structure, up to equivalence.
Consider, for instance, the triangle $T$ (not shown) in which 
$s_2 \rightarrow s_3$ by a shift, $s_2 \rightarrow s_1$ by base pair removal, 
and $s_1 \rightarrow s_3$ by a base pair addition. Then $T$ is equivalent to 
a triangle of type $A$.  Other instances of triangles the
reader may consider are analogously equivalent to a triangle of type 
$A$ or $B$.
}
\label{fig:trianglesAndTriples}
\end{figure}

\begin{figure}
\centering
\includegraphics[width=\textwidth]{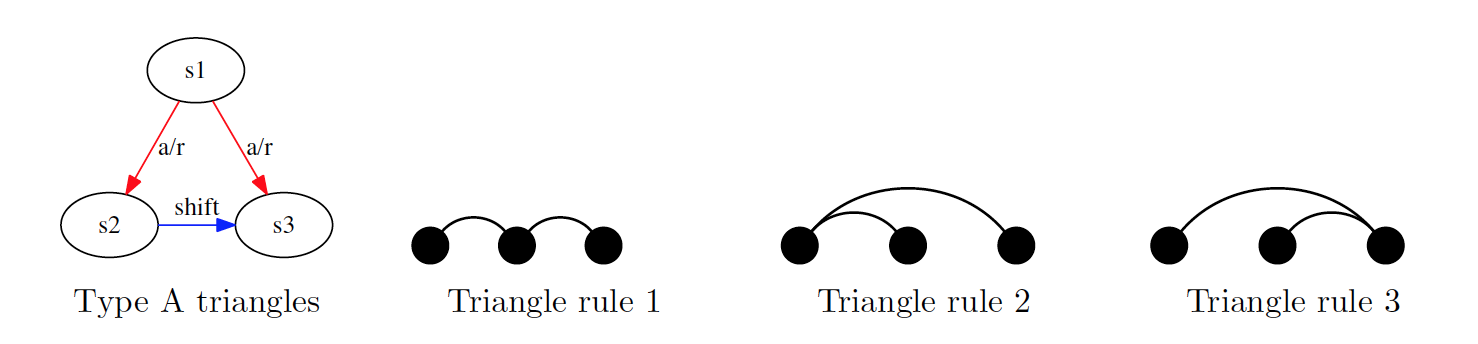}
\caption{Type A triangles are constituted by structures $s_1,s_2,s_3$,
where $s_2,s_3$ are obtained from $s_1$ by adding a base pair with the
property that there is a shift move from $s_2$ to $s_3$. This type of
triangle is described in rules 1,2,3 in Section~\ref{section:triangles}.
For instance, the motif for triangle rule 1
indicates that there is a base pair $(x,y) \in s_2$ that can be shifted
to the base pair $(y,z) \in s_3$; similarly, the motif for rule
2 [resp. 3]
indicates that there is a base pair $(x,y) \in s_2$ [resp. $(x,z)\in s_2$]
that can be shifted to the base pair $(x,z) \in s_3$ [resp. $(y,z) \in s_3$].
}
\label{fig:typeAtriangles}
\end{figure}

\begin{figure}
\centering
\includegraphics[width=\textwidth]{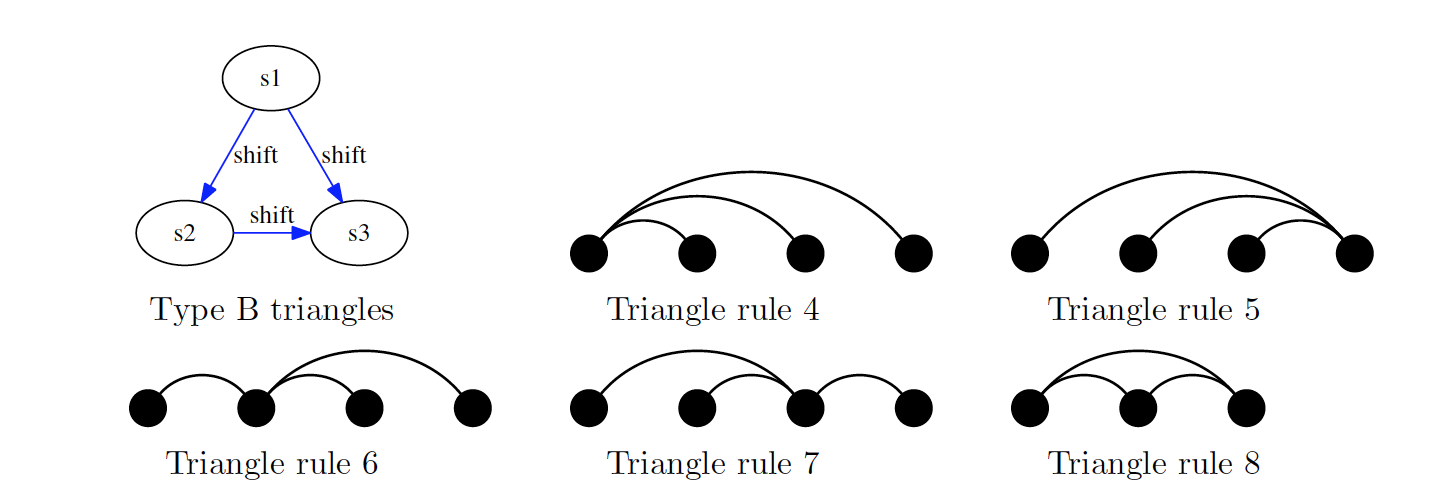}
\caption{Type B triangles are constituted by structures $s_1,s_2,s_3$,
where $s_2,s_3$ are obtained from $s_1$ by a shift with the
property that there is a shift move from $s_2$ to $s_3$. This type of
triangle is described in rules 4-8 in Section~\ref{section:triangles}.
For instance, the motif in triangle rule 4
indicates that there is a base pair $(x,y) \in s_1$ can be shifted to
base pair $(x,z) \in s_2$ and $(x,w) \in s_3$. The other panels have
analogous meanings.
}
\label{fig:typeBtriangles}
\end{figure}

\begin{figure}
\centering
\includegraphics[width=\textwidth]{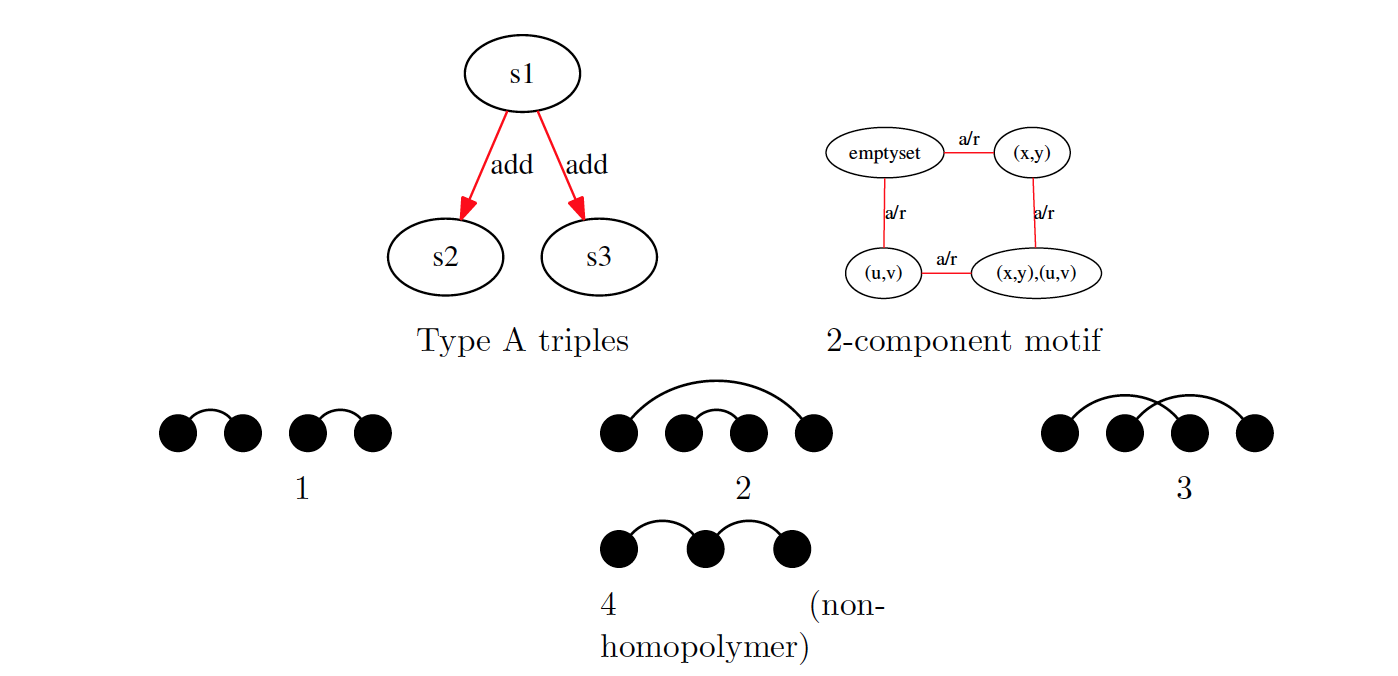}
\caption{Type A triples are constituted by structures $s_1,s_2,s_3$,
where $s_2,s_3$ are obtained from $s_1$ by a base pair addition with the
property that there is a no $MS_2$ move between $s_2$ and $s_3$. Type A
triples are given in rules 1-3 in Section~\ref{section:triples},
are are represented in panels 1,2,3,4 of this figure. Panel
1 [resp. 2] indicates that structures $s_2,s_3$ can be obtained from
structure $s_1$ by the addition of {\em disjoint} base pairs $(u,v)$ $(x,y)$
which are {\em not nested}, i.e. $\op \cp \op \cp$,
[resp. which are {\em nested}, i.e. $\op \op \cp \cp$].
Panel 3 indicates that structures $s_2,s_3$ can be obtained from
structure $s_1$ by the addition of {\em disjoint} base pairs $(u,v)$ $(x,y)$
which would form a pseudoknot if added simultaneously to $s_1$,
i.e. $\op \ob \cp \cb$. Panel 4, which is identical to
panel 1 of Figure~\ref{fig:typeAtriangles}, represents a non-triangular
connected triple {\em only} in the non-homopolymer case. This panel 
indicates that structures $s_2,s_3$ can be obtained from
structure $s_1$ by the addition of {\em non-disjoint} base pairs 
$(u,v)$ $(v,w)$ which share a base. To each triple motif that has
2 {\em extended connected components} (see text) there corresponds a
quadrilateral, as shown in the panel with label {\em 2-component motif},
where edges are labeled by {\em a/r} for base pair addition/removal.
To each corner of the quadrilateral, there corresponds a unique triple --
thus, panels 1 and 2 actually each represent 4 triples.
It follows that the average number of triples per structure for
type A(1) and type A(2) triples must be multiplied by 4.
The same remark holds for type C,D triples in
in Figure~\ref{fig:typeCDtriples}, which have 2 extended connected components.
}
\label{fig:typeAtriples}
\end{figure}

\begin{figure}
\centering
\includegraphics[width=\textwidth]{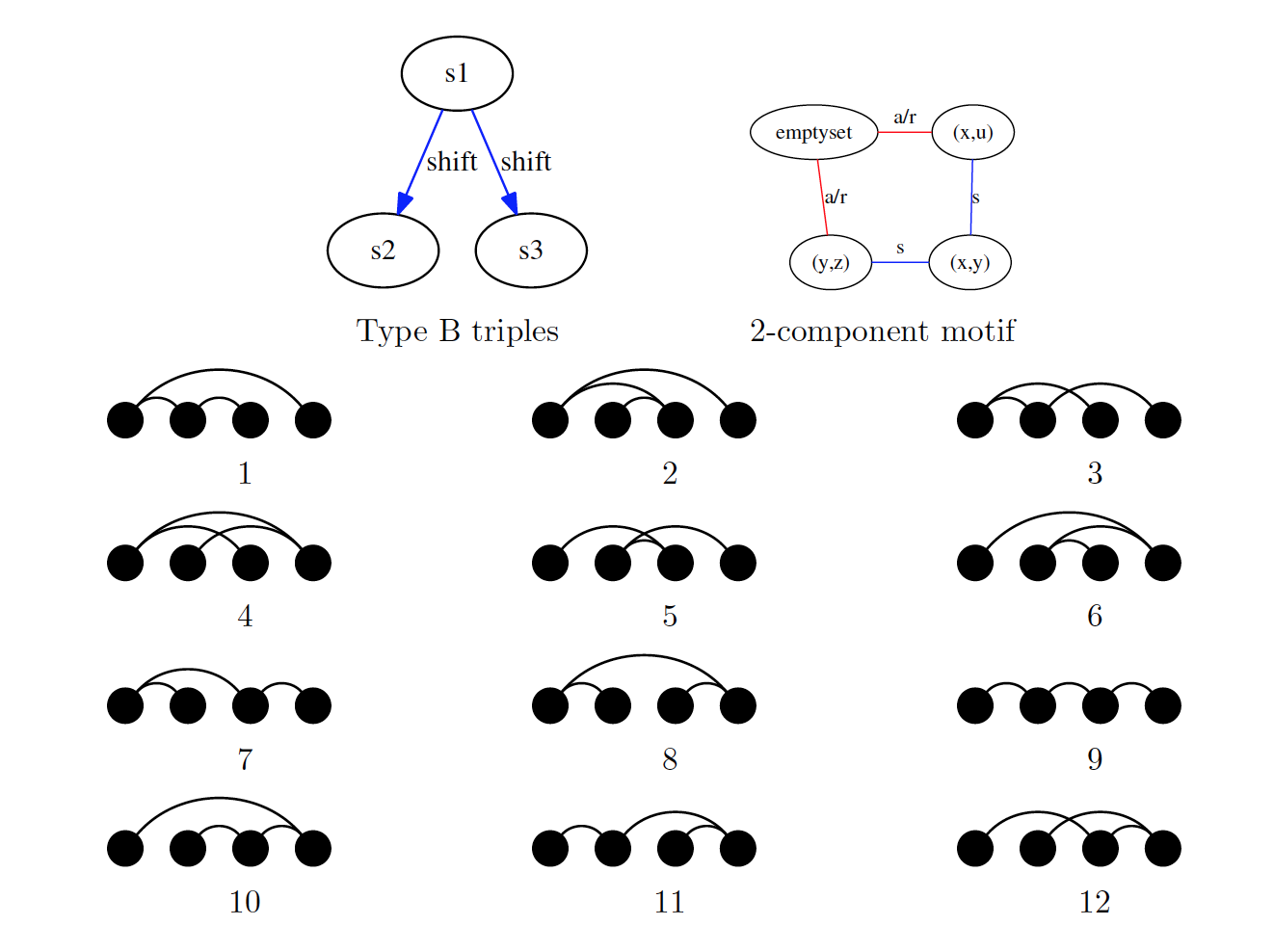}
\caption{Type B triples 
are of the form $s_1,s_2,s_3$, where $s_1,s_2$ are connected
by a shift, as are $s_1,s_3$, but $s_2,s_3$ are not connected by any
$MS_2$ move. Note that all motifs are connected.
To each type B triple motif, there corresponds a
quadrilateral, shown in panel with label {\em 2-component motif},
where edges are labeled by {\em a/r} for base pair addition/removal or
by {\em s} for base pair shift. For each of these 12 motifs, there is
a unique base pair that can shift to the remaining two base pairs -- for
instance, in motif 1, the base pairs are $(x,y)$, $(y,z)$ and $(x,u)$, 
for $x<y<z<u$, where
$(x,y)$ can be shifted to each of $(y,z)$ and $(x,u)$. This uniquely defined
base pair should be located in the quadrilateral diagonally opposite 
the corner labeled by {\em emptyset}. Since each corner of the quadrilateral
corresponds to one of 4 triples associated with the motif, it follows that
the average number of type B triples per structure must be multiplied by 4.
}
\label{fig:typeBtriples}
\end{figure}

\begin{figure}
\centering
\includegraphics[width=\textwidth]{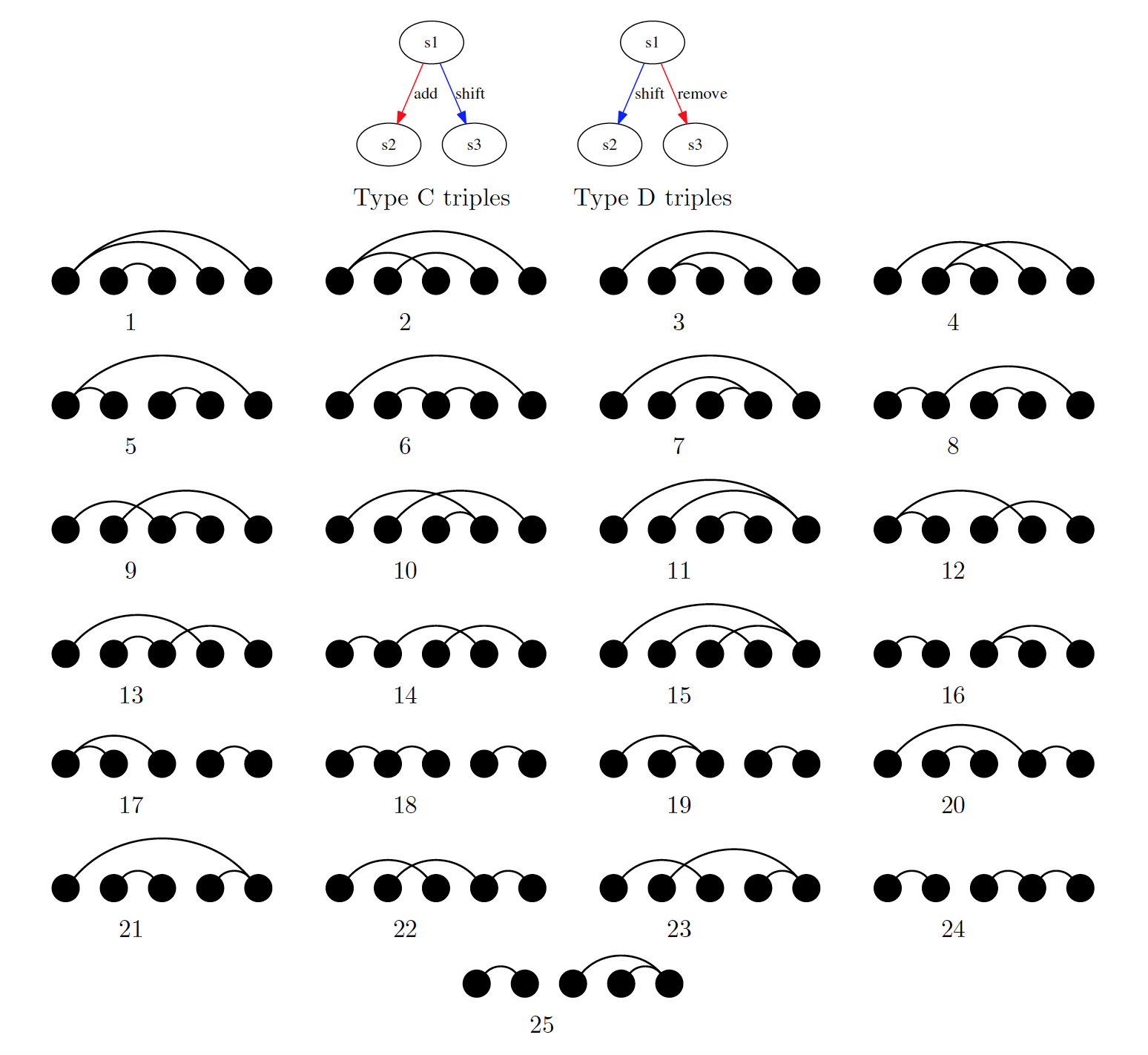}
\caption{Type C and D triples
are of the form $s_1,s_2,s_3$, where $s_1,s_2$ are connected
by a shift, and $s_3$ is obtained from $s_1$ by a base pair addition,
but $s_2,s_3$ are not connected by any $MS_2$ move. Triples may have either
one or two {\em extended connected components} (see text); 
in the former case, the asymptotic
expected number of triples is $O(n)$, while in the latter case the 
expected number of triples is $O(n^2)$, as is the case for the 
type A triple motifs in panels 1 and 2 of  Figure~\ref{fig:typeAtriples}. 
The 10 motifs having 1 extended connected component are:
2, 4, 9, 10, 12, 13, 14, 15, 22, 23.
The 15 motifs having 2 extended connected component are:
1, 3, 5, 6, 7, 8, 11, 16, 17, 18, 19, 20, 21, 24, 25. Of the latter,
the six motifs 16,17,18,19,24,25 are {\em disconnected successive}, and the
nine motifs 1,3,5,6,7,8,11,20,21 are {\em disconnected nested}.
To each motif, which has 2 extended connected components, there actually
correspond 4 triples, as explained in the caption to 
Figure~\ref{fig:typeAtriples}.
For instance, to the motif 16, given by undirected
edges $(1,2),(3,4),(3,5)$, there correspond four structures $s_1,s_2,s_3,s_4$,
where $(3,4) \in s_1$, $(3,5) \in s_2$, $(1,2),(3,4) \in s_3$,
$(1,2),(3,5) \in s_4$, with the following four triples:
(1) $s_1 \rightarrow s_2$, $s_1 \rightarrow s_3$ (type C triple);
(2) $s_2 \rightarrow s_3$, $s_2 \rightarrow s_1$ (type C triple);
(3) $s_3 \rightarrow s_4$, $s_3 \rightarrow s_1$ (type D triple);
(4) $s_4 \rightarrow s_3$, $s_4 \rightarrow s_2$ (type D triple).
This is analogous to the situation summarized in the panel in 
Figure~\ref{fig:typeAtriples} with label {\em 2-component motif}.
Note that motifs 3,6,7 have a type A triangle contained within an
outer designated base pair $\ob \cb$.
}
\label{fig:typeCDtriples}
\end{figure}